\documentclass[epj]{svjour}
\usepackage{graphics}
\usepackage{graphicx}
\begin{document}
%
\title{Collective strategy condensation towards
class-separated societies}
\subtitle{An envy-induced phase transition}
\author{Claudius Gros} 
%
%
\institute{Institute for Theoretical Physics, Goethe University Frankfurt, 
Frankfurt am Main, Germany}

\date{March 2022}
%
\abstract{
In physics, the wavefunctions of
bosonic particles collapse when the
system undergoes a Bose-Einstein
condensation. In game theory, the
strategy of an agent describes the
probability to engage in a certain course
of action. Strategies are expected to
differ in competitive situations, namely
when there is a penalty to do the same
as somebody else. We study what happens
when agents are interested how they fare
not only in absolute terms, but also
relative to others. This preference,
denoted envy, is shown to induce the
emergence of distinct social classes
via a collective strategy condensation
transition. Members of the lower class
pursue identical strategies, in analogy
to the Bose-Einstein condensation, with
the upper class remaining individualistic.
%
%
} 
\maketitle
%
\section{Introduction}

Humans do not live in isolation. It is well 
established that the social context is important, 
influencing, beside others, memory, cognition, 
risk awareness, accountability and decision making
\cite{wyer2014memory,linde2012social,lahno2015peer,tetlock1985accountability,tindale2019group}.
An important example is the emergence of cooperation, 
for which non-monetary components to the utility are
needed within a game-theoretical setting. Examples
are reputation seeking 
\cite{milinski2002reputation,hilbe2012emergence,kurokawa2009emergence}, 
the desire to minimize risks \cite{raub1997gains,hagel2016risk}, 
or the willingness to socialize \cite{szolnoki2016leaders}.

A generic non-monetary contribution to utility 
functions is a term arising from comparing rewards.
Here we follow \cite{gershman2014two} and use 
`envy' for contribution to game-theo\-re\-ti\-cal utility 
function that models inter-agent comparisons
\cite{alicke2008social}. Alternatively
one could speak of an aversion to unfairness.
Envy arises from the observation that the 
satisfaction individuals gain from making 
and spending money depends not only on absolute 
levels, but in good part also on how their own 
consumption level compares with that of
others \cite{hopkins2004running}. The
importance of comparing ourselves to others 
is also at the core of relative income theory 
\cite{mcbride2001relative,clark2010compares}.
Relative success is furthermore fundamental
for competitions in general, being it a track race
or evolutionary competition 
\cite{gros2015complex,hill2008evolutionary}.
 
For the study of envy, we examine games in 
which a psychological component is added to 
the classical, the monetary payoff 
\cite{capraro2021mathematical}. 
Our focus is the question whether
envy may have a qualitative impact on the 
structure of self-consistent multi-player
Nash states. We find that this is the case,
in the sense that that competitive societies
split endogenously into distinct social classes
when envy becomes relevant.

One-to-one inter-agent interactions are not present
in our framework. This implies, that the sole 
determinant for behavior are the individual payoffs,
which can be evaluated both in absolute and in 
relative terms. An analogy between agents 
and the distinguishable particles of
classical physics can hence be made 
\cite{gros2021collective}, with the game-theoretical
space of available options corresponding to
the physical state space. Strategy condensation
occurs when the initially distinct strategies of 
a macroscopically large number of agents collapses 
into a single, encompassing strategy. This process
can be considered as a classical correspondence
of the Bose-Einstein condensation (BEC), which is
characterized by the macroscopic occupation of
a single state, usually the ground state
\cite{griffin1996bose}.  At its core, BEC is 
however a quantum phenomenon. A similar analogy 
in term of a classical de-facto 
Bose-Einstein distribution has been pointed out 
in the context of complex networks \cite{bianconi2001bose}.

\section{Framework}

In a first step we discuss our approach
to model a society of competing agents,
adding subsequently the psychological 
component, envy.

Agents have $i\in[1,N]$ possible options,
with each options yielding a basic payoff, 
$v_i$. This is the monetary income the
agent receives if nobody else would
select the same option. For the modeling
of the inter-agent competition, we define
with $p_i^\alpha\ge0$ the strategy of agent 
$\alpha\in[1,M]$, namely the probability to select 
option $i$, with the normalization 
condition $\sum_i p_i^\alpha=1$.
The number of options and agents is 
respectively $N$ and $M$. Strategies 
are pure when $p_i^\alpha$ reduces to 
a discrete $\delta$-function, as
$p_i^\alpha\to\delta_{ik}$, being 
mixed otherwise. The real-world income 
$I^\alpha$ of an agent is given by
\begin{equation}
I^\alpha = \sum_i I_i^\alpha p_i^\alpha,
\qquad\quad
I_i^\alpha = v_i\left(1-\kappa\sum_{\beta\ne\alpha} p_i^\beta \right)\,,
\label{I_alpha}
\end{equation}
where $\kappa\ge0$ encodes the strength of the
competition. Payoff reduction is proportional 
to $\sum_{\beta\ne\alpha} p_i^\beta$, which
corresponds in lowest order (modulo higher-order
combinatorial factors) to the probability
to encounter other agents. Overcrowded 
options yielding negative payoffs will 
be avoi\-ded. In this study we set $\kappa=1/2$, 
which describes the case that two agents with 
identical pure strategies share resources 
equally. More than two agents can opt for 
the same course of action $i$ hence only 
when playing mixed strategies, 
with $p_i^\alpha<1$.

The exact functional form of the bare 
utility $v_i$ is of minor importance. 
Mapping options $i$ to a scalar quantity, 
the quality $q_i\in[0,1]$, allows to 
define a functional representation for
$v_i=v(q_i)$. We use
\begin{equation}
v(q_i) = \frac{1+q_i}{1-\theta q_i},
\qquad\quad
\theta\in[0,1[\,,
\label{v_q}
\end{equation}
with the parameter $\theta$ regulating the
relative width of the distribution of 
bare utilities. One has $v(q_i)\in[1,2/(1-\theta)]$,
which translate into
$v(q_i)\in[1,4]$ when $\theta=0.5$.

\subsection{Nash equilibria without envy}

It is assumed throughout this study that the
basic utility function $v_i$ is non-generate, 
viz that $v_i\ne v_j$ for all option pairs 
$i\ne j$. 
Per se, without an additional psychological 
component, the utility defined by (\ref{I_alpha}) 
has then a straightforward solution. Agents 
play exclusively pure strategies, with either one or 
two agents per option. Throughout this study we 
denote these two types of pure strategies as 
`pure-1' and `pure-2'.

\subsection{Information requirement}

Formulating the competition $\sim\!\kappa$ 
in terms of strategies, as done in
(\ref{I_alpha}), demands at face sight
a high amount of information, 
namely the knowledge of the strategies
$\rho_i^\alpha$ of all participating 
agents. This is however not the case.
For a proof we rewrite the probability 
to encounter other agents at a given
option $i$ as
\begin{equation}
\sum_{\beta\ne\alpha} p_i^\beta = 
\sum_{\beta} p_i^\beta - p_i^\alpha\,,
\label{kappa_mean_field}
\end{equation}
which is solely a function of the agent's
own strategy, $p_i^\alpha$, and the overall
mean number $N_i=\sum_{\beta} p_i^\beta$ of 
agents selecting the option in question.
Agents need to be aware hence only of their
own strategy and of the mean occupations $N_i$,
but not of the specifics of the 
strategies of everybody else.

\subsection{Envy -- comparing success}

A basic reference level for success 
within a well-mixed population is 
the average income $\bar{I}$,
\begin{equation}
\bar{I} = \frac{1}{M}\sum_\alpha I^\alpha\,,
\label{bar_I}
\end{equation}
with $I^\alpha$ being defined by 
(\ref{I_alpha}). On the level of 
individuals, envy results from direct
person-to-person comparison. Here
we examine instead the population effect,
for which the average income $\bar{I}$ 
is taken as the yardstick. Both approaches,
using $\bar{I}$ as a reference, or
pairwise comparisons, become identical,
as shown in the appendix, when income 
differences are small.

In terms of the overall reward function 
$R_i^\alpha$, the effect of envy is encoded by
\begin{equation}
R_i^\alpha = I_i^\alpha + 
\epsilon\, p_i^\alpha
\log\left(\frac{I^\alpha}{\bar{I}}\right)\,,
\label{R_i_alpha}
\end{equation}
where the first term on the right-hand side
corresponds to the monetary utility. The
strength of second term, the psychological 
component, is regulated by $\epsilon\ge0$. 

The rationale for the functional form of the
envy term in (\ref{R_i_alpha}) is straightforward.
The agent is happy when earning above average,
when $I^\alpha>\bar{I}$, and unhappy otherwise.
The $\log$-dependency ensures that the envy term
contributes positively in the first, and negatively
in the second case. The envy term couples directly
to the strategy, $p_i^\alpha$, with the consequence
that the current strategy is enforced when 
$I^\alpha>\bar{I}$. Below the average, when
$I^\alpha<\bar{I}$, agents are motivated in
contrast to search for alternatives. It is 
also worth mentioning that the 
log-dependency $\log(I^\alpha/\bar{I})$ of the
envy term in (\ref{R_i_alpha}) is in agreement with
the Weber-Fechner law, which states that the brain 
discounts sensory stimuli \cite{hecht1924visual},
numbers \cite{dehaene2003neural},
time \cite{howard2018memory}, and
data sizes \cite{gros2012neuropsychological}
logarithmically. 

In previous studies envy has been assumed 
to result from comparing rewards 
\cite{gros2020self,gros2021collective},
and not incomes, as done here. In that
case the envy term involves
$\log(R^\alpha/\bar R)$, instead of
(\ref{R_i_alpha}), where $\bar R$ is 
the mean reward. The resulting framework
demands however that agents can estimate
the psychological state of the other agents,
which is included in $R^\alpha$, but not 
in $I^\alpha$, which seems to be less 
plausible. A well-defined large-size
limit is attained in any case for constant 
$\epsilon$ when scaling $M$ and $N$ such that 
the ratio $\nu=M/N$, the fraction 
of agents per options, is retained \cite{gros2020self}.

\subsection{Numerics}

Agents maximize their expected 
rewards $R^\alpha$,
\begin{equation}
R^\alpha = \sum_i R_i^\alpha p_i^\alpha\,
\label{bar_R}
\end{equation}
viz the sum of the monetary utility 
$I^\alpha$ and of the envy term. Numerically,
this is achieved using standard evolutionary 
dynamics \cite{hofbauer2003evolutionary},
\begin{equation}
p_i^\alpha(t+1) = \frac{p_i^\alpha(t)R_i^\alpha(t)}
{\sum_j p_j^\alpha(t)R_j^\alpha(t)}\,,
\label{evolutionaryDynamics}
\end{equation}
which describes the transition of strategies 
at time $t$ to $t+1$. Adding a constant 
offset $R_0$ to the bare rewards $R_i^\alpha$ 
leads to a smooth convergence. If not 
otherwise stated, $R_0=20$ has been used.

Once (\ref{evolutionaryDynamics}) has been
iterated till self-consistency, the resulting
payoffs $R_i^\alpha$ are positive. When negative
values appear as an intermediate step, they
are set to a lower bound $R_0$, which has been
taken to be zero. It is advantageous to use 
the representation (\ref{kappa_mean_field})
for the interaction term. We focused here on
the case $M=N$, together with $\kappa=0.5$ and
$\theta=0.5$. Testing extensively for alternative
parameter settings we find that the overall 
picture emerging is remarkably robust.

For the results presented below the evolution dynamics
(\ref{evolutionaryDynamics}) is iterated $10^5$ times, 
if not otherwise stated. The initial strategies 
$p_i^\alpha(t\!=\!0)$ are drawn randomly from $[0,1]$, 
and normalized subsequently. We did test that the 
resulting strategy distributions are stable against 
small stochastic perturbations. The resulting states
are hence `local' Nash-equilibria, in the sense that
it is disadvantageous for agents to make small changes
to their $p_i^\alpha$.

\begin{figure}[!t]
\centering
\includegraphics[width=0.85\columnwidth]{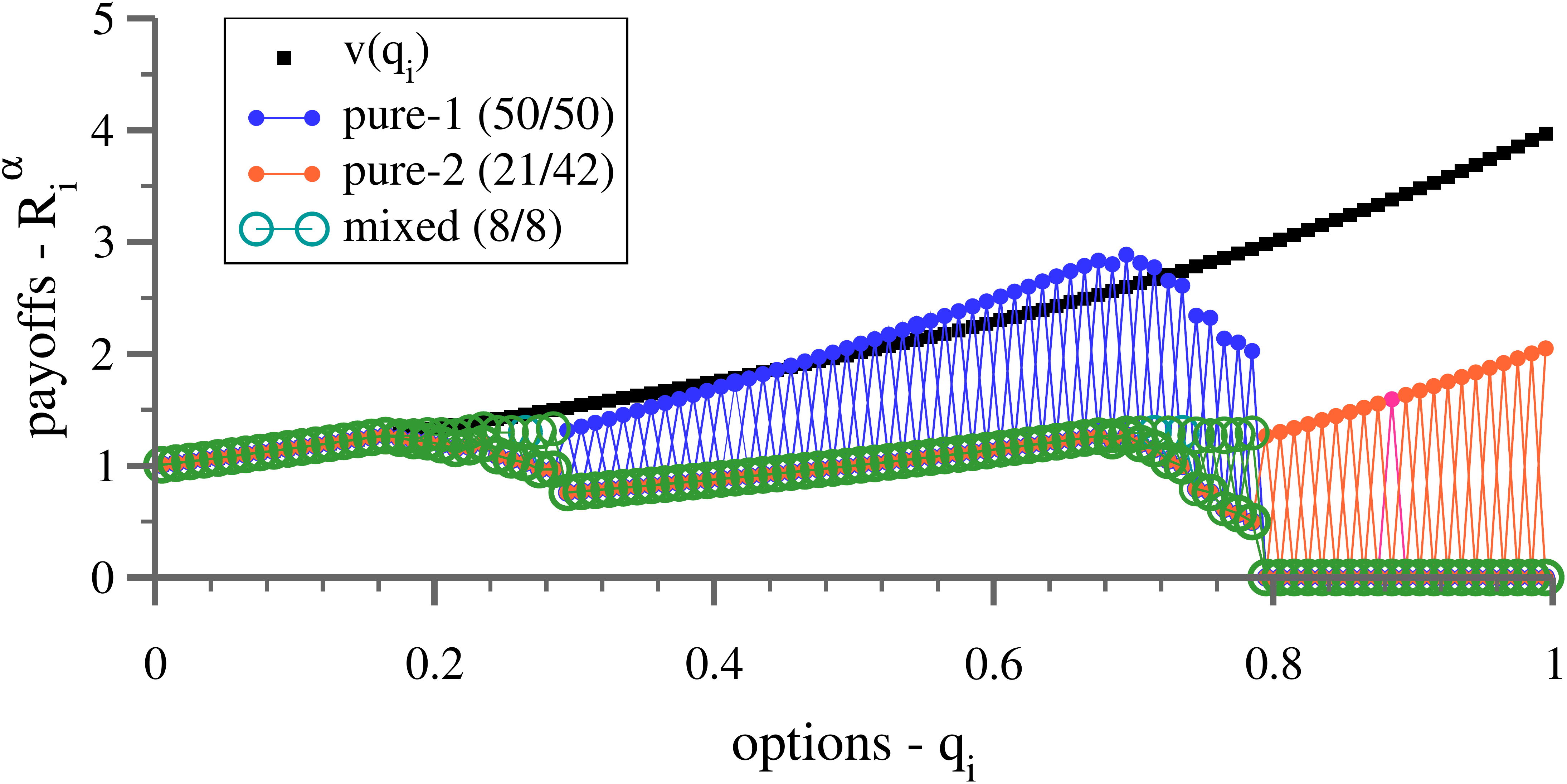}
\caption{{\bf Assorted competition.} The 
payoffs of $\alpha\in[1,100]$ agents (circles), 
$i\in[1,100]$ options, and $\epsilon=1$. The
data points of the payoff functions $R_i^\alpha$ 
are connected by lines. 50 agents engage in pure 
strategies which are not contested (pure-1, blue), 
with $21$ agents playing pure strategies that 
are pursued also by another agent (pure-2, red). 
The rest, $8$ agents, follow distinct mixed 
strategies (green), with an average support of $2.9$.
Also given is the bare utility $v(q_i)$ (black).
Shown in the brackets, $(N_s/N_p)$, is the
number $N_s$ of distinct strategies of a given 
type, together with the number $N_p$ of 
engaged agents.}
\label{fig_payoffs_100_rel_e10}
\end{figure}

\section{Results}

We analyze our results in particular with 
respect to the size of the support $S^\alpha$ 
of the individual strategies $p_i^\alpha$,
\begin{equation}
S^\alpha = \big\{i\,|\,p_i^\alpha>0\big\}\,,
\label{S_alpha}
\end{equation}
which is given by the set of options selected 
with finite probability. The smallest possible 
support is one, corresponding to the case of a 
pure strategy. Strategies with non-trivial
support are mixed.

\subsection{Low envy: assorted competitive state}

In Fig.~\ref{fig_payoffs_100_rel_e10} a typical
payoff configuration for $\epsilon=1$ is given. 
Shown are the entirety of all $M=100$ payoff 
functions $R_i^\alpha$, viz the respective
values for all $i=1,..,N$ options, as defined by
(\ref{R_i_alpha}). Larger bare utilities 
$v(q_i)$ are taken each by two agents playing 
pure strategies, a configuration denoted 
with `pure-2'. This region cannot be invaded 
by competitors, given that $I_i^\alpha$, as 
defined by (\ref{I_alpha}), vanishes 
in the pure-2 region  for third parties.
Most agents are engaged in `pure-1' strategies,
which means that they play pure strategies that
are contested, if at all, only by agents 
playing mixed strategies, but not by other 
agents pursing pure strategies. 

In the absence of envy, when $\epsilon=0$, only
pure-1 and pure-2 strategies are present. For small
values of envy, as for the data presented in
Fig.~\ref{fig_payoffs_100_rel_e10}, mixed strategies
start to invade the pure-1 region at the boundary
to the pure-2 strategies, with additional support 
at low values of $v(q_i)$. Mixed strategies are
distinct for low $\epsilon$, which means that
no two agents pursue identical mixed strategies.
Supports are comparatively small. Overall, a 
competitive Nash equilibrium with assorted strategies 
is observed.

\begin{figure}[!t]
\centering
\includegraphics[width=0.85\columnwidth]{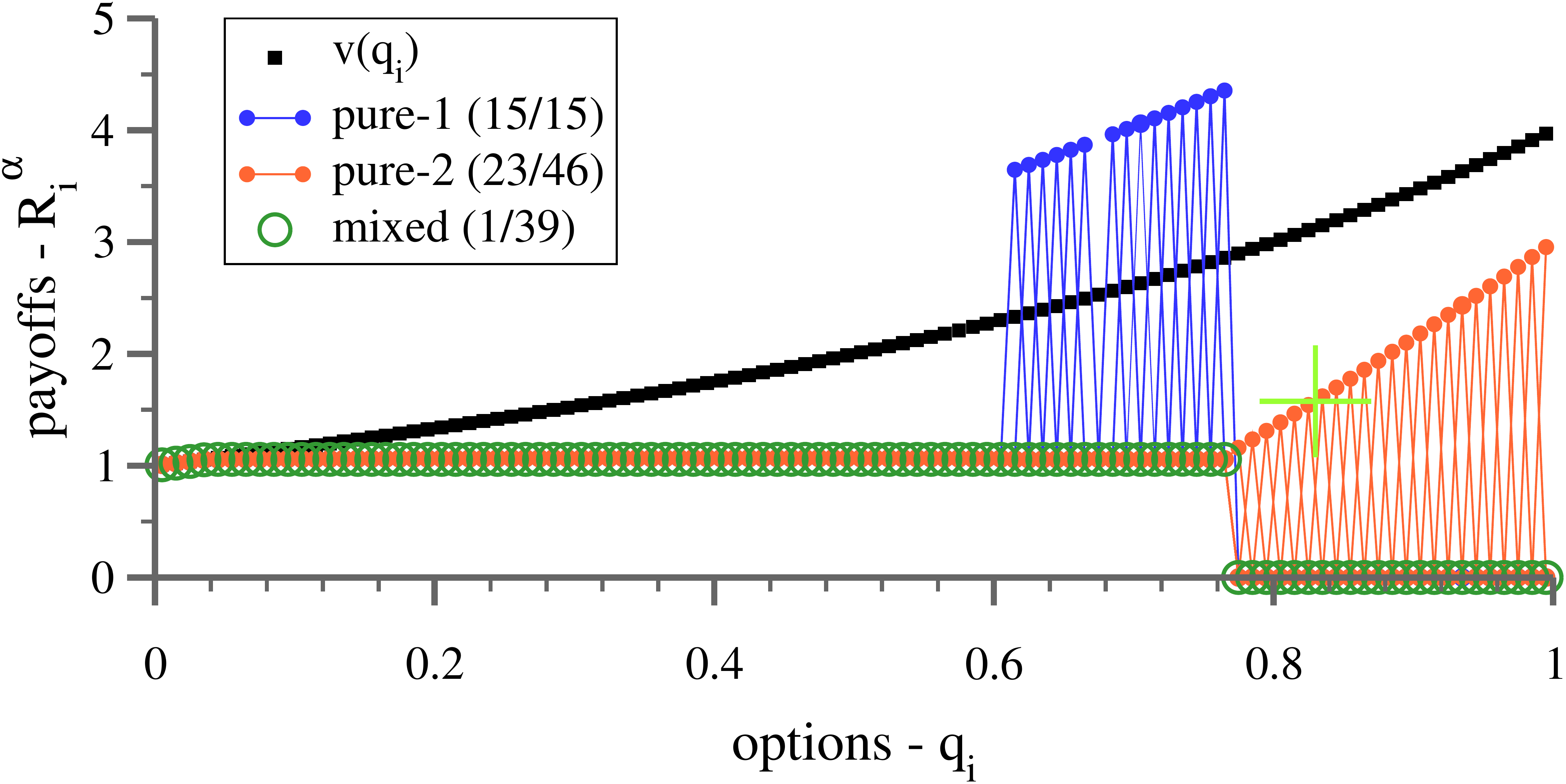}
\caption{{\bf Class-stratified state.}
The payoffs of $M=100$ agents (circles), for 
$N=100$ options $q_i$ and $\epsilon=4$. The
lines connect payoffs $R_i^\alpha$ for one 
and the same agent $\alpha$. There
are $15$ agents playing pure strategies which
are not contested (pure-1, blue), with $23$ agents 
playing pure agents that are played also by
another agent (pure-2, red). The rest, $39$
agents, play the identical mixed strategy (green)
covering $62$ options.
Also shown is the bare utility $v(q_i)$ (black).
The cross (light yellow) indicates the locus of 
vanishing envy, viz where $I^\alpha=\bar{I}$.
The brackets, $(N_s/N_p)$, show the number $N_s$ 
of distinct strategies of a given type together 
with the number of agents $N_p$ playing the 
respective type of strategy.
}
\label{fig_payoffs_100_rel_e40}
\end{figure}

\subsection{Large envy: condensed mixed strategies}

The spectrum of payoffs functions reorganizes 
dramatically for larger value of $\epsilon$. 
Not with regard to the two sets of pure-1 and 
pure-2 strategies, which change only with
respect to their sizes, but regarding the 
mixed strategies. All mixed strategies collapse 
into a single strategy pursued by a finite 
fraction of agents, the lower class. The 
situation is illustrated in 
Fig.~\ref{fig_payoffs_100_rel_e40} for
$\epsilon=4$. About 40\% of all agents
play one and the same mixed strategy, with
the respective support covering about 60\% 
of option space.

The details of the individual strategies are 
private information within the framework 
examined here. Only the cumulative weights,
$\sum_\beta p_i^\beta$, are available,
see (\ref{kappa_mean_field}). There is
consequently no inbuilt mechanism allowing 
an agent to copy somebody else's strategy. 
The incomes received, $I^\alpha$, are
however public, which means that they can
be compared with each other, the core 
functionality of envy. The condensation
of the strategies of a large number of agents 
is therefore a collective phenomenon in
the sense of statistical physics 
\cite{gros2015complex}. It occurs when
increased levels of envy lead to a substantial
inter-agent coupling. 

Agents are identical, apart from their
strategies, which are drawn at the start
from a flat random distribution. Chance, 
the starting strategy, determines therefore
the outcome -- whether the agent will belong
in the end to the individualistic
pure-1 or pure-2 clusters, or whether it will
become part of the masses, loosely speaking,
sharing the same mixed strategy with many
others.

\subsection{Social classes separated by income gaps}

\begin{figure}[!t]
\centerline{
\includegraphics[width=0.75\columnwidth]{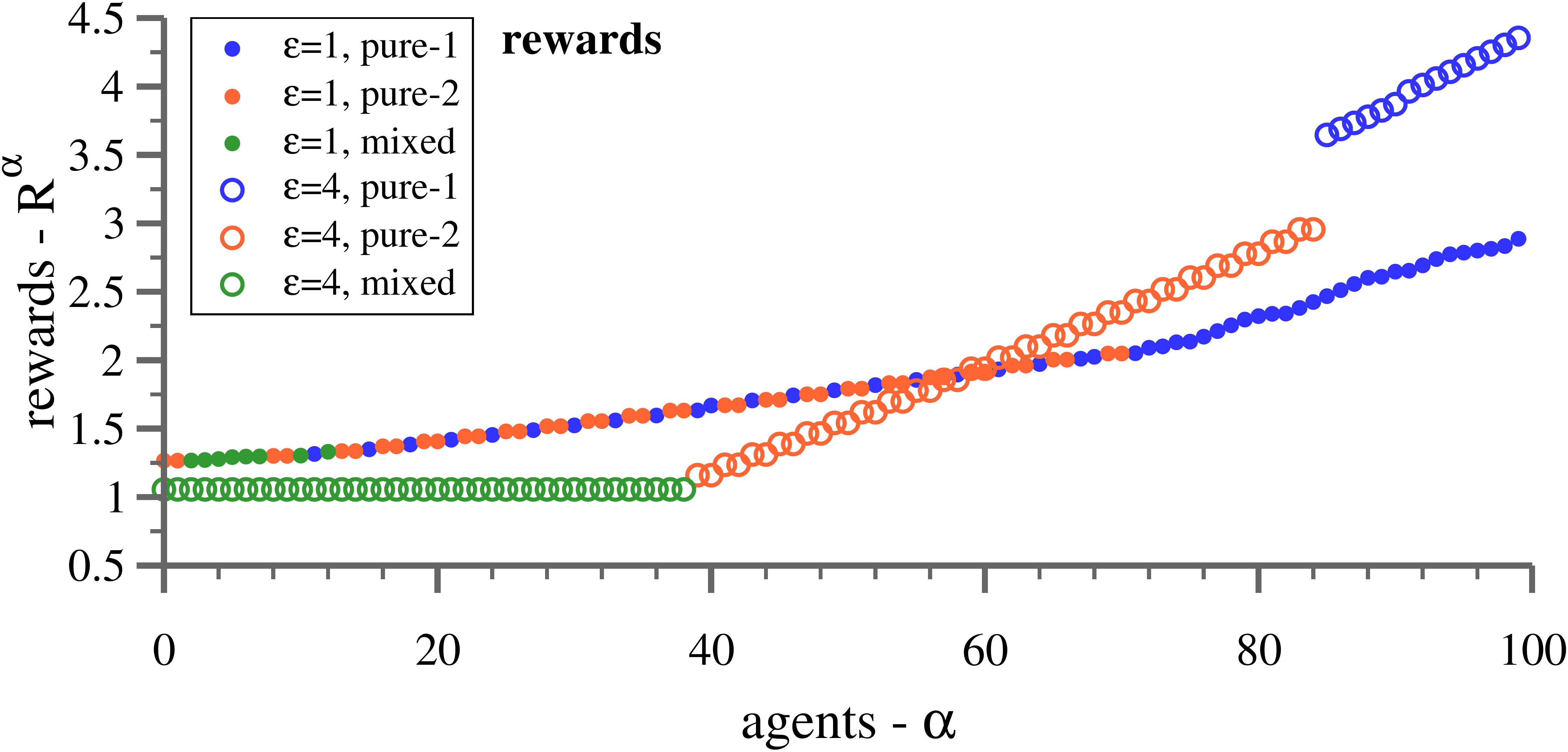}
           }
\centerline{
\includegraphics[width=0.75\columnwidth]{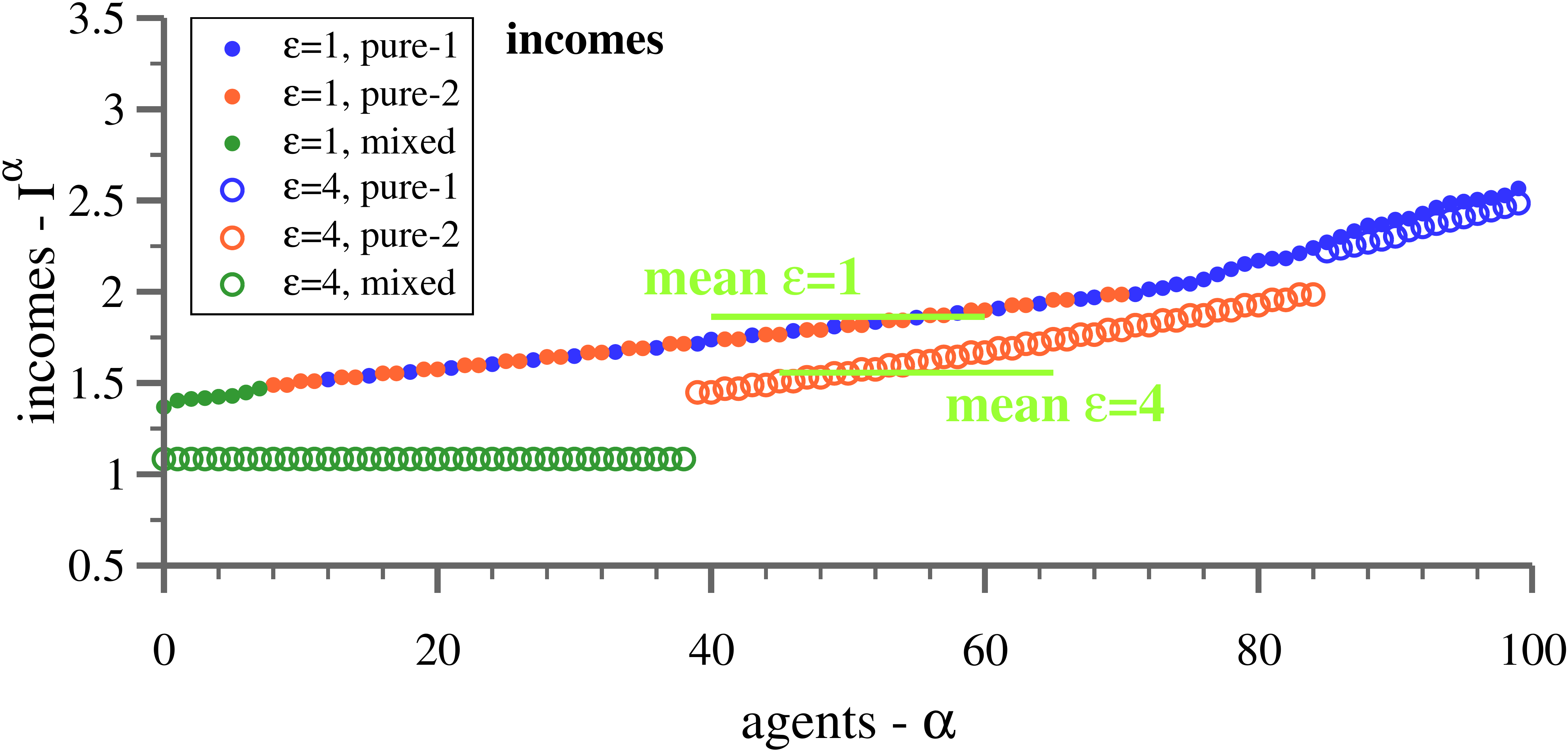}
           }
\caption{{\bf Player specific rewards and incomes.}
For all agents, the rewards $R^\alpha$ (top panel) 
and the incomes $I^\alpha$ (bottom panel), as sorted by value.
Shown are the results corresponding to the payoffs 
presented in Figs.~\ref{fig_payoffs_100_rel_e10} and
\ref{fig_payoffs_100_rel_e40}, for both
$\epsilon=1$ (filled symbols) and
$\epsilon=4$ (open symbols). Also shown are
the population averages $\bar{I}$ for the incomes
(light green). Note the income gaps for 
$\epsilon=4$.
}
\label{fig_rewards_incomes_e10_e40}
\end{figure}

In Fig.~\ref{fig_rewards_incomes_e10_e40} we
present the player specific averages
corresponding to the payoff functions shown 
in Figs.~\ref{fig_payoffs_100_rel_e10} and
\ref{fig_payoffs_100_rel_e40}. For pure
strategies, the rewards $R^\alpha$ correspond to
the respective peaks of the payoff functions.
When sorted by ascending values, the rewards
of pure-1, pure-2 and mixed strategies are
intertwined for $\epsilon=1$, but not for
$\epsilon=4$. It is also noticeable that
there is no gap in the $\epsilon=4$
reward spectrum between mixed and pure-2
strategies, whereas a substantial gap
is present for the corresponding income
data. This phenomenon can be traced back
to a discontinuity of the player specific
envy $E^\alpha$,
\begin{equation}
E^\alpha = R^\alpha-I^\alpha = \epsilon\log\left(
\frac{I^\alpha}{\bar{I}} \right)P_2^\alpha,
\qquad
P_2^\alpha=\sum_i \big(p_i^\alpha\big)^2\,,
\label{E_alpha}
\end{equation}
compare (\ref{R_i_alpha}). For pure strategies
$P_2^\alpha\to1$ holds, but not for 
mixed strategies. This is true in particular 
for the mixed strategy of the condensed state 
realized for $\epsilon=4$, which has a substantial 
support and correspondingly small probabilities 
$p_i^\alpha$. The envy term $E^\alpha$ jumps 
hence together with $P_2^\alpha$ at the
boundary of the mixed and the pure-2 cluster,
which have respectively low/large values 
of $|E^\alpha|$. Note that $I^\alpha<\bar{I}$ 
at the boundary, as indicated in 
Fig.~\ref{fig_payoffs_100_rel_e40}, which 
implies a negative $E^\alpha<0$ and that
pure-2 strategies must have larger incomes
at boundary to the lower class. The latter
due to the continuity of the reward 
spectrum $R^\alpha$.

The occurrence of substantial gaps in the income
spectrum shows that the society of agents separates 
spontaneously into distinct social classes under 
the influence of envy \cite{gros2021collective}. 
Incomes are distributed over a finite range
also for low values of envy, as evident from 
the data presented in 
Fig.~\ref{fig_rewards_incomes_e10_e40}. The
resulting distributions are however continuous, 
which implies that there is no unbiased criterion 
to subdivide the society into separated social 
classes. The situation changes beyond the
strategy condensation transition, where three
differentiated strategy clusters emerge:
\begin{itemize}
\item {\bf Lower class:} A finite fraction of agents
(`the masses') pursues the identical mixed strategy,
with incomes that are below the average, $I^\alpha<\bar{I}$.

\item {\bf Middle class:} The middle class claims
the options with the highest bare utilities
$v(q_i)$. The price paid is that options are selected
by pairs of agents, which halves the respective
monetary yields.
The resulting incomes $I^\alpha$ are both below and
above the population average $\bar{I}$.

\item {\bf Upper class:} Agents playing pure strategies
that are contested only by lower-class agents, but not
by other pure strategies.
\end{itemize}
The here used definition of social classes 
in terms of separated income clusters is 
convenient for modeling studies, but less 
suitable for demographic investigations. 
Real-world income distributions 
do not show clear income gaps \cite{yakovenko2007two}.

\begin{figure}[!t]
\centerline{
\includegraphics[width=0.75\columnwidth]{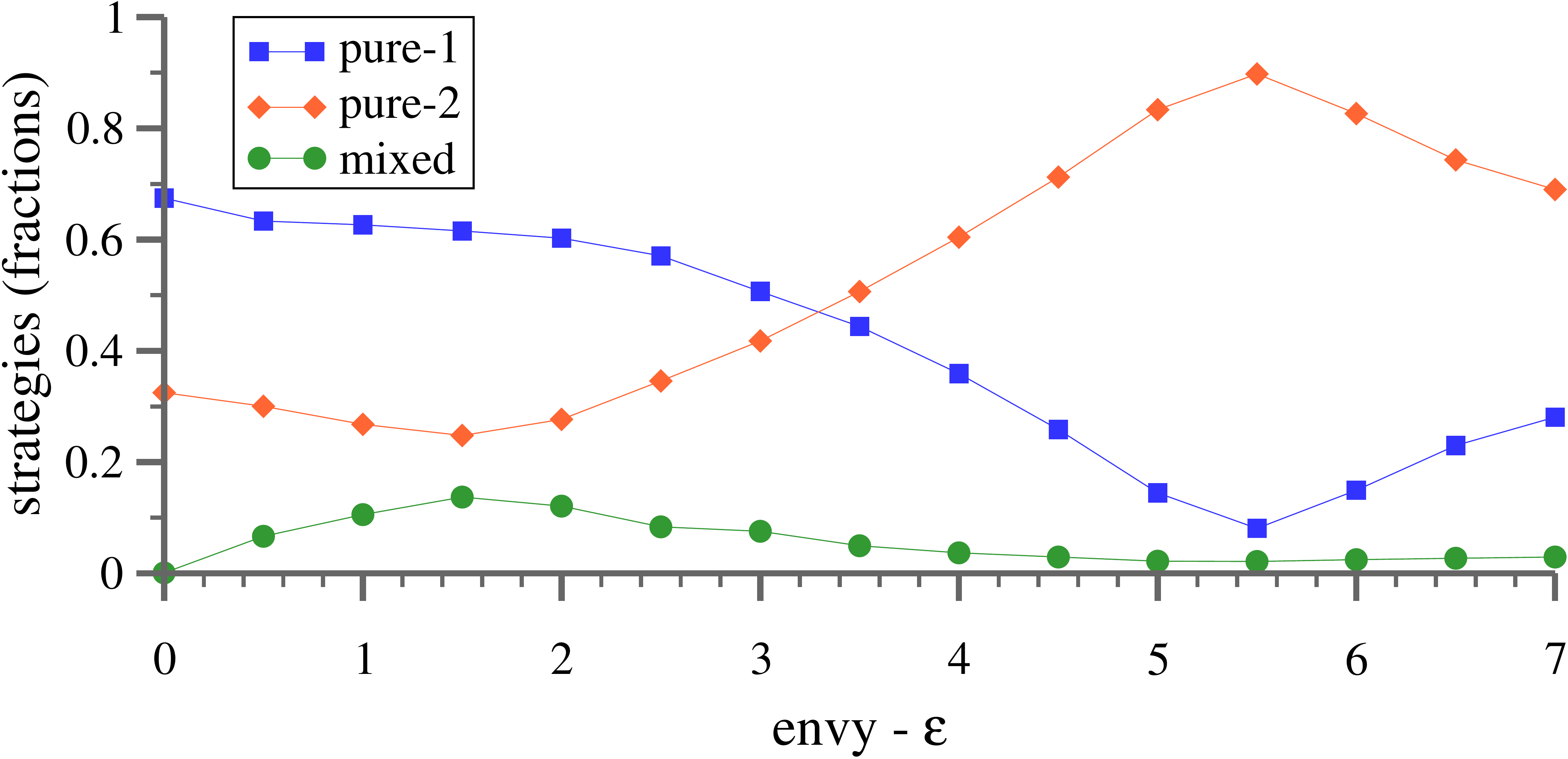}
           }
\centerline{
\includegraphics[width=0.75\columnwidth]{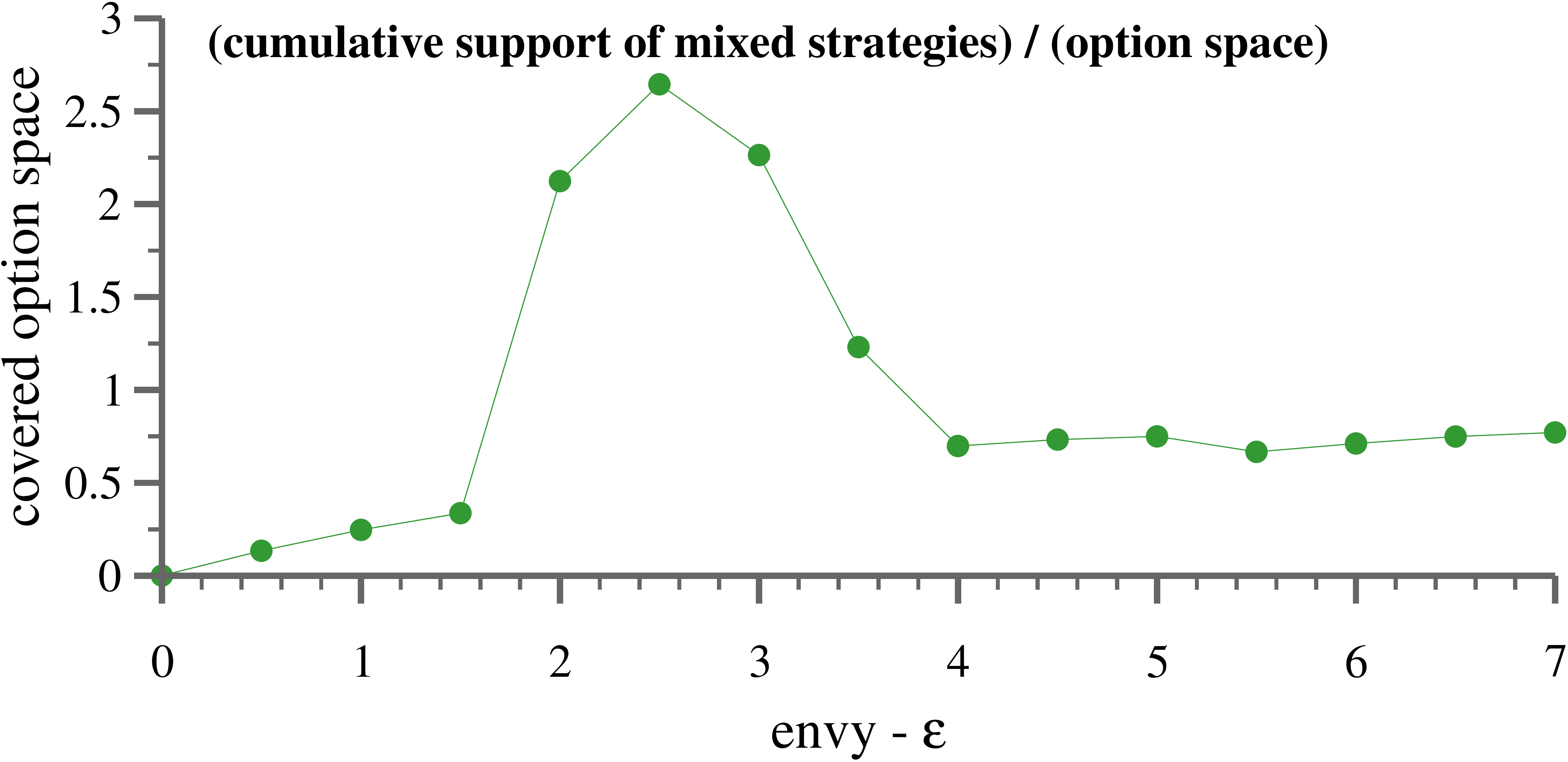}
           }
\caption{{\bf Evolution of strategies played.}
{\bf Top:} For $N=M=200$ the fractions of observed 
strategies, as a function of envy. Shown are the 
relative fractions of different strategies, for
pure-1 (blue), pure-2 (red) and mixed (green). 
Counted are the numbers of strategies, not agents. 
The data is averaged over ten random initial conditions.
{\bf Bottom:} The cumulative support of all mixed
strategies divided by the number of available options,
$N$. The ratio can exceed unity, as mixed strategies 
may partially overlap.
}
\label{fig_envy_strategies_rel}
\end{figure}

\subsection{Strategy condensation transition}

In Fig.~\ref{fig_envy_strategies_rel}
the evolution of key strategy-related quantities
is shown as a function of envy. Data for slightly
larger numbers of agents and options,
$N=200=M$, has been averaged over ten random 
initial conditions. The absolute system size
is of minor importance, as long as the ratio
$M/N$ of agents per options is kept
\cite{gros2020self}. The fractions of strategies 
played are given relative to the total number
of observed distinct strategies. For the Nash equilibrium
of Fig.~\ref{fig_payoffs_100_rel_e10}, to give
an example, there are $50+21+8=79$ different
strategies. The fractions of pure-1, pure-2 and
mixed strategies would then be $50/79$, $21/79$ 
and $8/79$. One observes that mixed strategies
are most important, in terms of relative numbers,
around $\epsilon\approx1.5$.

Beyond $\epsilon\approx5.5$, the mixed strategy 
of the lower-class starts to invade intermittently 
the cluster of pure-2 strategies, which shrinks 
consequently in terms of relative importance. 
The strict ordering of the supports of the mixed 
and the pure-2 strategies in $q_i$-space, 
as presented in Fig.~\ref{fig_payoffs_100_rel_e40}, 
is hence lost for very large values of $\epsilon$.

Also included in
Fig.~\ref{fig_envy_strategies_rel}
is the total support of all mixed strategies,
with mixed strategies played by more than one
agent counted only once. The total support
has been divided by the overall number of 
options, $N$, with values above unity indicating
that mixed strategies overlap on the average.
The option space covered jumps substantially
between $\epsilon=1.5$ and $\epsilon=2.0$,
due to the emergence of mixed strategies with
extended supports. Between $\epsilon=1.5$ and 
$\epsilon=2.0$ the average relative support 
of mixed strategies rises eight-fold, from 
$3.0$ to $23.9$.
Total support drops beyond the peak, at 
$\epsilon\approx 2.5$, due to corresponding 
decrease in the number of mixed strategies.
At this stage there is typically one large
condensed mixed strategy, together with
a few smaller mixed strategies, all leading
to very similar incomes. The number of 
smaller mixed strategies becomes progressively
smaller with increasing envy. For $\epsilon=4$
and $\epsilon=6$ there are, e.g., on the average
$2.9$ and respectively $1.0$ mixed strategies,
with actual numbers fluctuating between individual 
simulations and initial conditions. In
Fig.~\ref{fig_payoffs_100_rel_e40} the case
of a single mixed strategy has been shown
for $\epsilon=4$. Residual smaller mixed strategies
would have been essentially indistinguishable
on the scale.

It is important to point out, that the 
strategy condensation transition appears
from replicator dynamics, see
Eq.~(\ref{evolutionaryDynamics}), which is 
the only way for agents to evolve their
strategies. Other forms of learning, f.i.\ by 
joining other strategies, are not present.
The formation of an extended class of agents
with (numerically) identical strategies is
hence due to a collective effect.

\begin{figure}[!t]
\centerline{
\includegraphics[width=0.75\columnwidth]{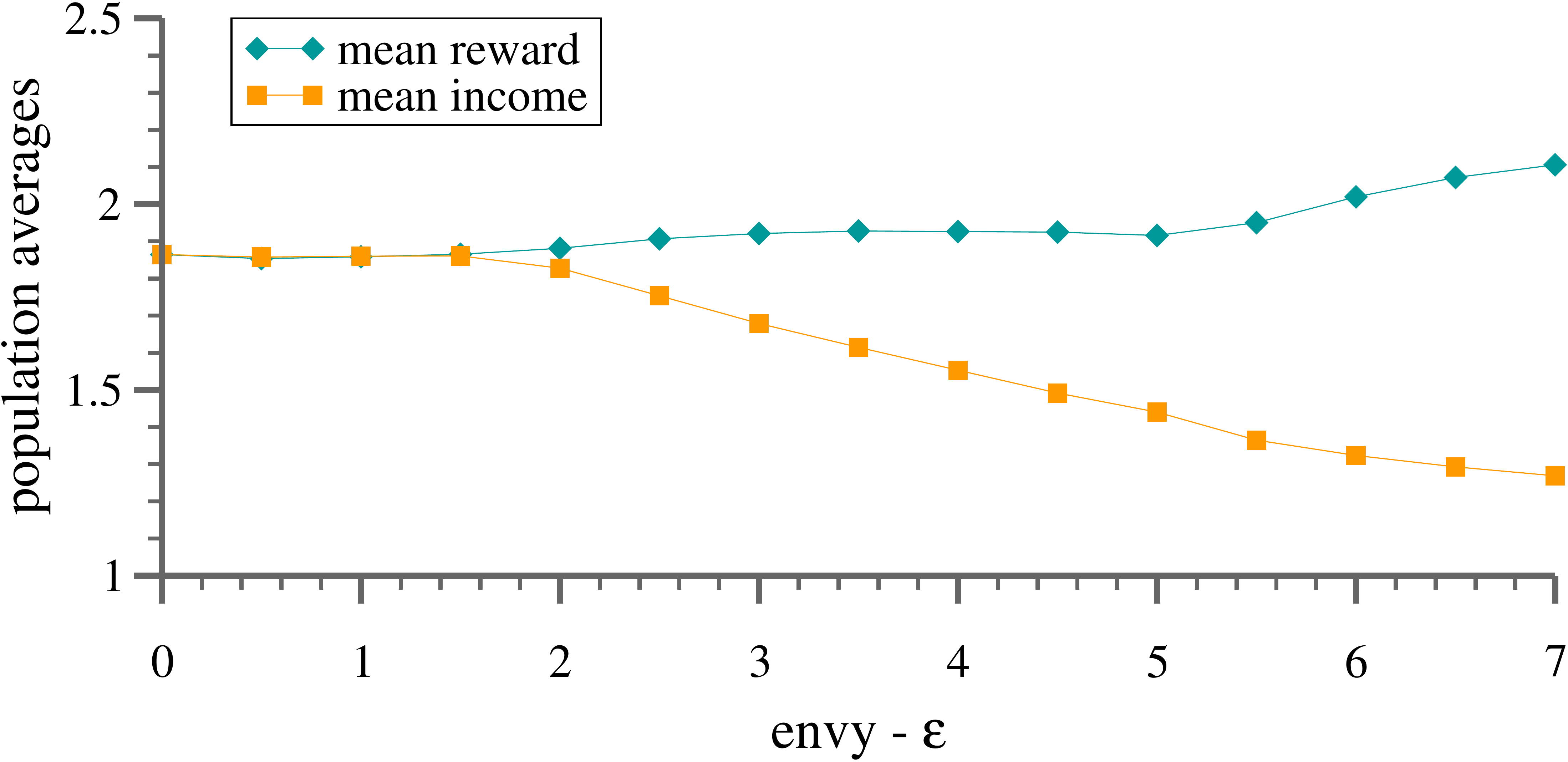}
           }
\caption{{\bf Envy reduces monetary incomes.}
Mean incomes $\bar{I}$ and rewards, $\bar{R}$,
as defined by (\ref{bar_I}) and (\ref{bar_R}).
As a function of envy, mean incomes drop substantially
beyond the strategy condensation transition, which
starts between $\epsilon=1.5$ and $\epsilon=2$.
Envy, a psychology component, affects real-world
monetary incomes adversely.
}
\label{fig_envy_rewardsIncomes_rel}
\end{figure}

\subsection{Social dilemma}

In Fig.~\ref{fig_envy_rewardsIncomes_rel}
the evolution of the population averages 
$\bar{I}$ and $\bar{R}$ with envy are 
presented, respectively for incomes and 
rewards. The mean reward varies comparatively
little with $\epsilon$, overall an increase
from $1.86$ to $2.11$ can be observed between
$\epsilon=0$ to $\epsilon=7$. The situation
is different for $\bar{I}$, which drops 
noticeably within the class-stratified state.
Once strategy condensation sets in, around
$\epsilon=1.5-2.5$, incomes start to drop 
progressively.

The evolutionary benefits of envy as a 
multi-faceted human emotion are not fully 
understood \cite{ramachandran2017evolutionary}. 
It is however generically assumed that envy
is a direct consequence of the core defining 
feature of Darwinian evolution, namely that 
only the `fittest' survives. To be among 
the top performers is defined in relative, 
and not in absolute terms 
\cite{hill2008evolutionary,garay2011envy}.
Our results, that envy is detrimental to the
overall welfare of societies, suggest in this
perspective a social dilemma. As an operative 
human psychological trait 
\cite{alicke2008social}, envy seems to have
disparate effects on individual and societal
levels. It is hence important to study, as done
in the following sections, what happens when 
envy is not a uniform characteristic, but
a heterogeneous, player-specific trait.

\subsection{Player-specific envy levels}

The stability of the  phenomenology emerging 
from our framework, that a strategy 
condensation transition is induced by raising 
levels of envy, can be studied against various
perturbations. Here we examine the case that
agents have distinct degrees of envy.
For this question we use a particularly 
straightforward protocol, namely that the
player specific envy-parameters $\epsilon^\alpha$
are uniformly distributed on the interval
$[0,2\bar{\epsilon}]$, where $\bar{\epsilon}$
is the mean envy level. 

\begin{figure}[!t]
\centerline{
\includegraphics[width=0.75\columnwidth]{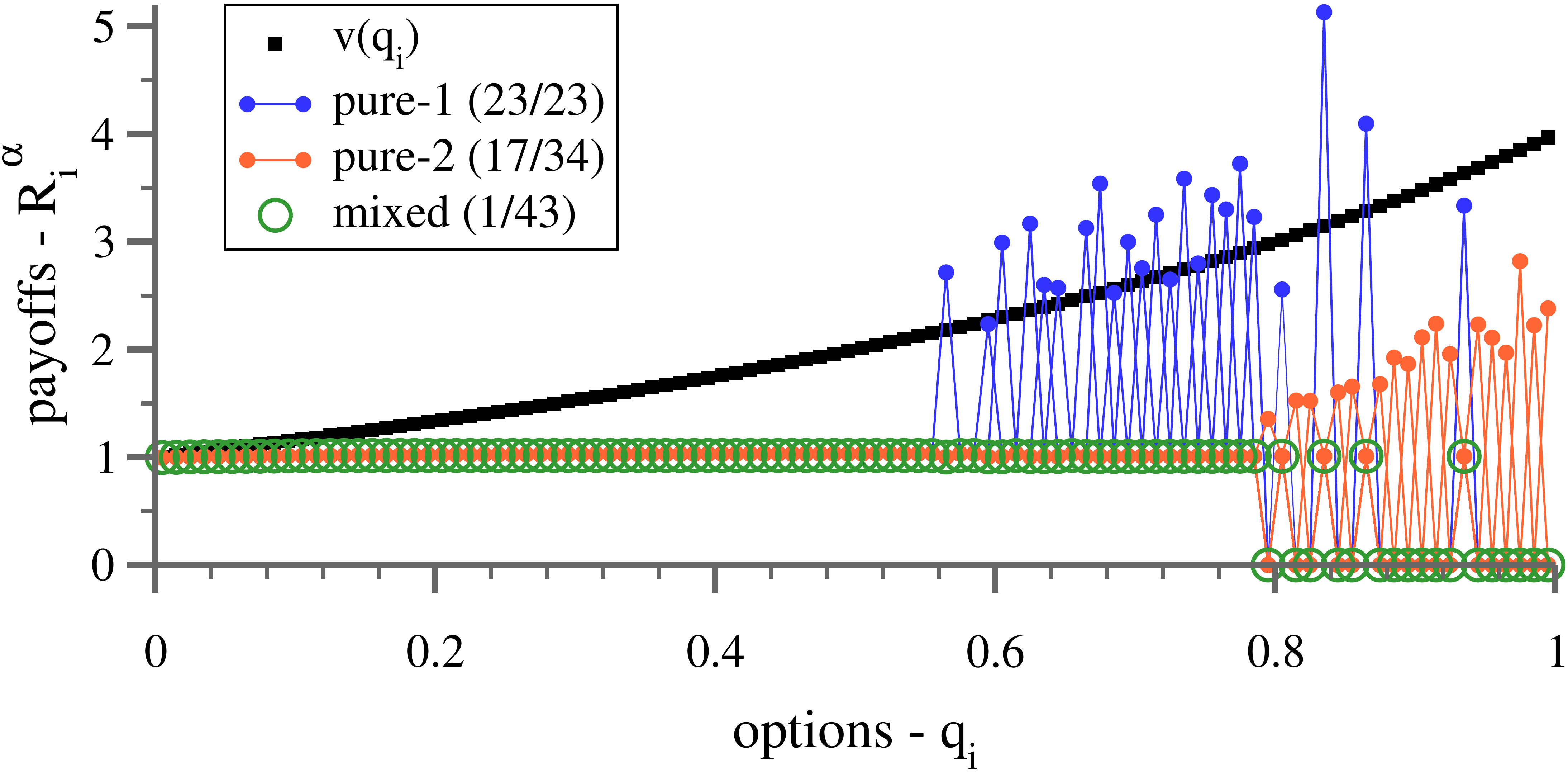}
           }
\centerline{
\includegraphics[width=0.75\columnwidth]{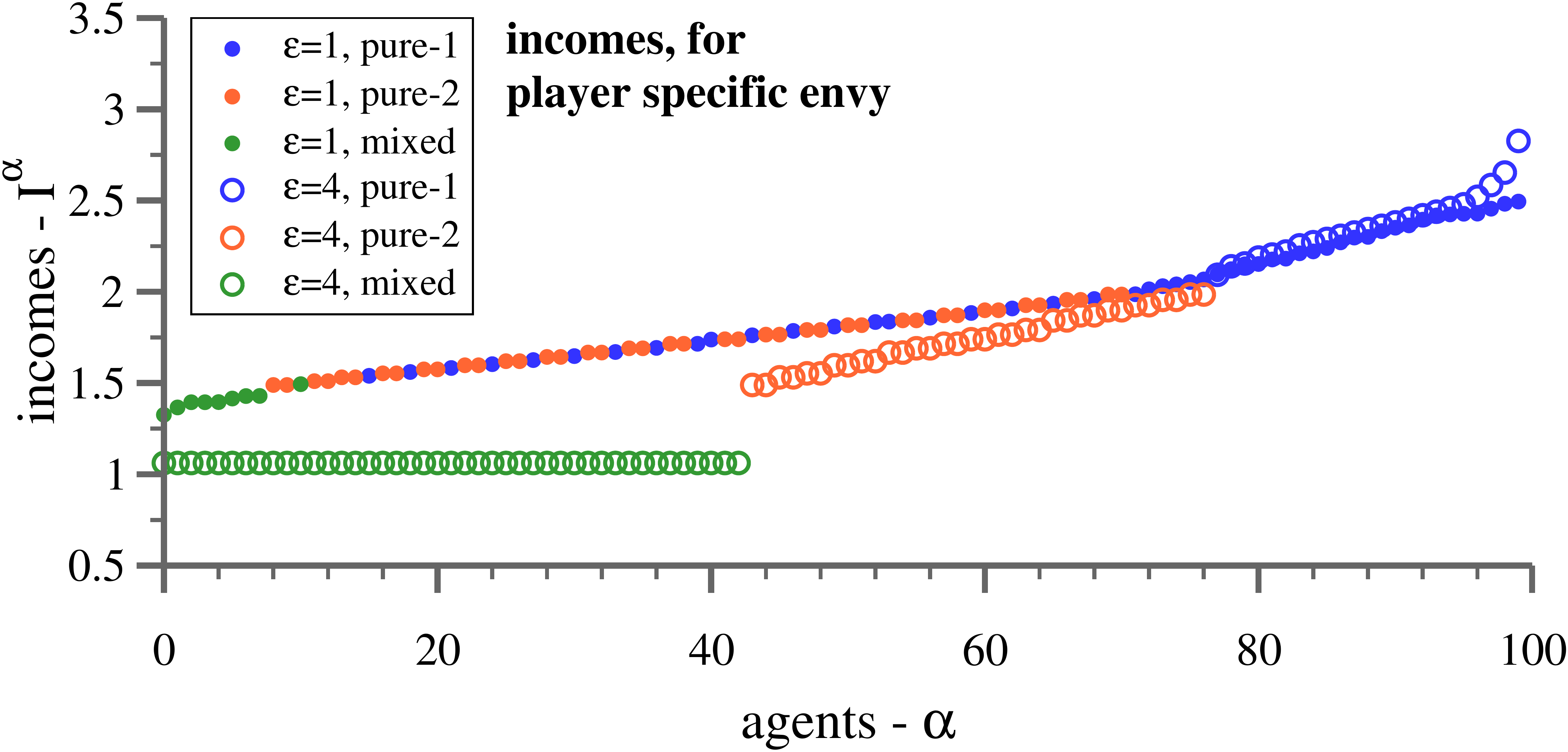}
           }
\caption{{\bf Agents with personalized levels of envy.}
{\bf Top:} As for Fig.~\ref{fig_payoffs_100_rel_e40},
but for agents with player-specific envies 
$\epsilon^\alpha$ that are distributed equally on the 
interval $[0,2\bar{\epsilon}]$, with a mean $\bar{\epsilon}=4$.
The `scattering' of the payouts for pure strategies is
a consequence of the varying levels of envy. Denoting
the mixed state (green) as (1/43), as done the legend, 
is an approximation. On a fine numerical level, the 
lower class is a (43/43) state, albeit with only 
very small payoff differences.
{\bf Bottom:} The respective average incomes, 
as in Fig.~\ref{fig_rewards_incomes_e10_e40}. 
Both for $\bar{\epsilon}=1$ and $\bar{\epsilon}=4$
(indicated by $\epsilon$ in the legend).
}
\label{fig_payoffs_100_bar_e40}
\end{figure}

One finds that varying agent specific levels
lead only to quantitative changes, leaving the
basic processes intact. A typical outcome 
for $N=M=100$ is shown in Fig.~\ref{fig_payoffs_100_bar_e40},
where the payoff functions are given for
$\bar{\epsilon}=4$, together with the ordered 
incomes $I^\alpha$. The latter is given
for comparison also for $\bar{\epsilon}=1$. 
Two observations are retained, the occurrence 
of a transition resembling the condensation of
strategies, and that a substantial income gap 
opens between the lower class and the 
rest of the society.
We notice that the mean income 
drops with raising $\bar{\epsilon}$,
here from $\bar{I}=1.85$ to $\bar{I}=1.58$ 
when going from $\bar{\epsilon}=1$
to $\bar{\epsilon}=4$. A social dilemma is 
hence present also when agents are characterized
not by uniform, but by player-specific levels of envy.

For $\bar{\epsilon}=1$ one finds
pure-1 (49/49), pure-2 (21/42) and mixed (9/9)
strategies. For $\bar{\epsilon}=4$ we have
pure-1 (23/23), pure-2 (17/34) and ``mixed (1/43)''
strategies. Beyond the transition, a lower
class emerges. At the resolution of the plot
presented in Fig.~\ref{fig_payoffs_100_bar_e40},
differences in respective rewards or incomes, 
$R^\alpha$ and $I^\alpha$, of the $43$ lower-class
agents cannot be discerned. It is however not 
possible that agents with different envy $\epsilon^\alpha$
satisfy (\ref{R_i_alpha}) for identical 
strategies and incomes. On a fine scale it should
hence be the case that the lower class splits into 
a mixed (43/43) state, albeit with only small 
differential payoffs. We verified that this concurs 
with the numerical results shown in 
Fig.~\ref{fig_payoffs_100_bar_e40} for 
$\bar{\epsilon}=4$, for which the lower 
class has a small but finite bandwidth.

It is remarkable, that the strict condensation 
of strategies observed for identical 
$\epsilon^\alpha\equiv\bar{\epsilon}$, is
retained to a very good approximation when envy
becomes specific to the individual agents.
For an understanding we recall that the expectation 
value (\ref{E_alpha}) of the envy term involves
with $P_2^\alpha=\sum_i (p_i^\alpha)^2$
the sum of the squared strategy. Strategies
with macroscopic support obey the scaling
$|S^\alpha|\sim N$, which implies 
$P_2^\alpha \sim N/N^2\sim 1/N$. In this
case, the envy term will vanish in the 
thermodynamic limit $N\to\infty$, with
differences in the $\epsilon^\alpha$ 
becoming irrelevant. The band of 
lower-class agents showing up in 
Fig.~\ref{fig_payoffs_100_bar_e40} will
hence become degenerate in the thermodynamic
limit.


\subsection{Correlations between income levels and envy}

\begin{figure}[!t]
\centerline{
\includegraphics[width=0.75\columnwidth]{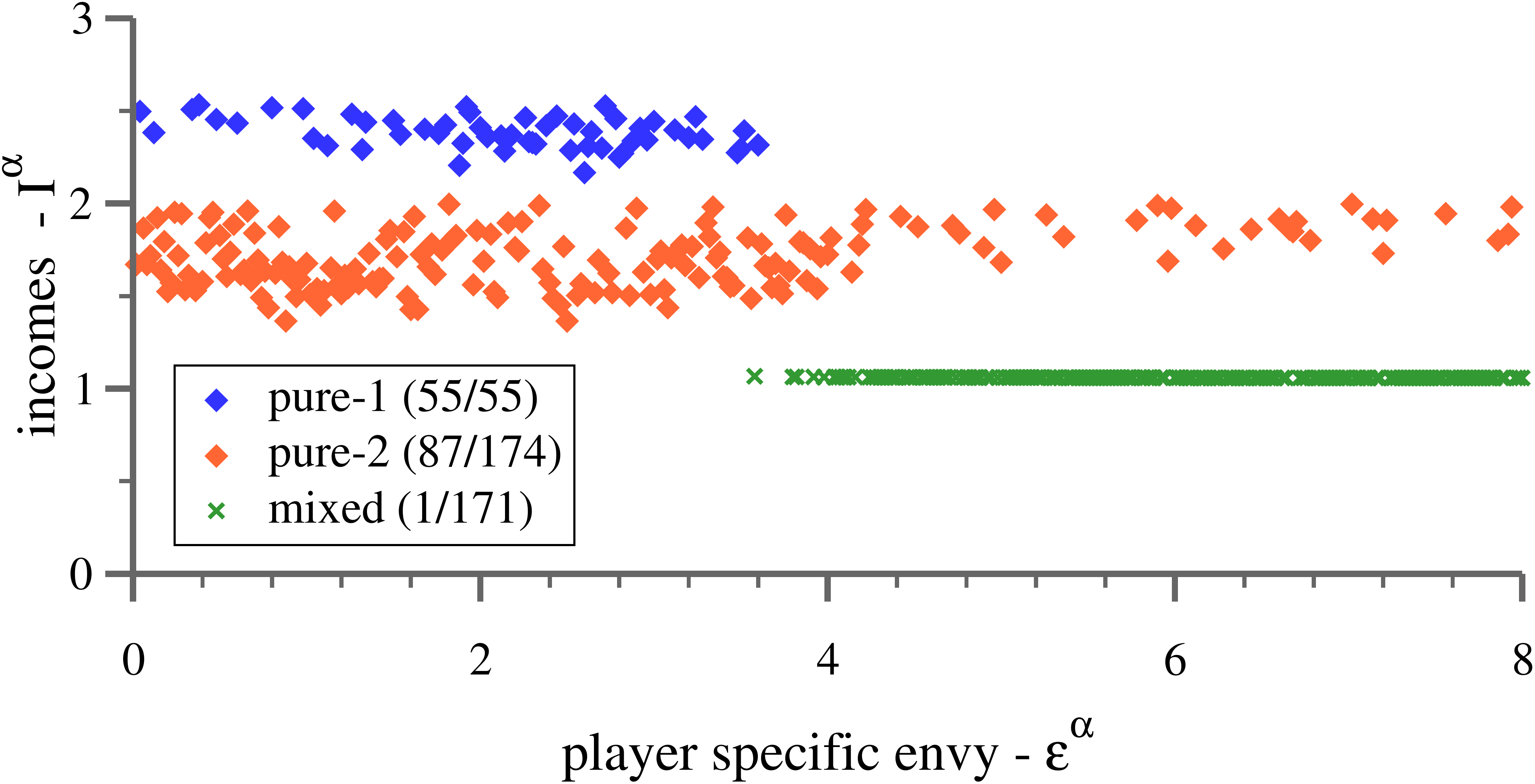}
           }
\caption{{\bf Envy specific incomes.}
For distributed player specific values of
envy, $\epsilon^\alpha\in[0,8]$, the incomes 
$I^\alpha$ for a society with $M=N=400$ 
agents and options. The numbers of agents
playing pure-1, pure-2 and mixed strategies
are $55$, $174$, and $171$, which adds to $M=400$. 
As for the data shown in Fig.~\ref{fig_payoffs_100_bar_e40},
the denotation (1/171) is an approximation.
On a fine numerical scale, the lower class (green) 
is composed of 171 distinct levels, being
a (171/171) state. Lower-class agents play very
similar, but not identical strategies. The three 
different types of strategies form non-overlapping 
clusters in the $(\epsilon^\alpha, I^\alpha)$ plane.
}
\label{fig_envy_scatter_bar}
\end{figure}

We investigated whether the player specific 
incomes $I^\alpha$ are correlated with
the respective envy-strengths, $\epsilon^\alpha$.
We used $\bar{\epsilon}=4$, as in
Fig.~\ref{fig_payoffs_100_bar_e40}, but this
time for a larger system, with $M=N=400$,
which allows for a reliable statistics. 
In Fig.~\ref{fig_envy_scatter_bar}
the results are shown as a scatter plot in
the $(\epsilon^\alpha, I^\alpha)$ plane.
It is clear that the pure-1, pure-2 and
mixed strategies form strategy-specific 
bands that are separated in the
$(\epsilon^\alpha, I^\alpha)$ plane.

The incomes $I^\alpha$ of lower class 
members range from 1.058 to 1.067,
which is just above the minimum of the 
bare utility, $v(0)=1$. The width of
the lower-class band of about 1\% is
proportional to sum over squared
strategies, as discussed above, and
hence small. The data presented in 
Fig.~\ref{fig_envy_scatter_bar} shows
furthermore that agents with low values 
of envy $\epsilon^\alpha$ can be found 
with a somewhat increased probability 
close to the mean income, $\bar{I}=1.525$, 
which was to be expected. Members of the 
lower class have in contrast substantial 
values of $\epsilon^\alpha$. When envy
is large, the incentive to search for
alternatives is substantial for agents
with below the average incomes.

\subsection{Ultimatum game}

Envy is a multi-faceted human trait
that may act both on societal and
on individual levels, with both 
types being equivalent to lowest 
order, as discussed in the appendix.
The latter observation allows to gauge 
the magnitude of $\epsilon$ from 
lab studies, which are available
for two-person games.

A reference game revolving around the
notion of fairness is the ultimatum game. 
\cite{thaler1988anomalies,kirchsteiger1994role}.
In the ultimatum game a given amount of
monetary utility $s_0$ is partitioned according 
to what one of the players, the proposer, 
suggests. Say as 70\%-30\% shares. The other 
player, the responder, may accept or decline. 
When declining, nobody gets anything. The 
expected outcome depends on whether the 
game is interpreted within a closed- or 
an open-room framework. Standard testing 
protocols presume the first case, namely
that the game takes place in a closed room 
separated from everything else. It would
then be rational for the responder to
accept offers of every size, even then
getting only 10\%, which is however not 
observed.

Experimentally, people tend to reject
offers that are too unfair. This makes 
sense in an open-room setting, namely 
for the case that both participants 
could potentially be competitors in the 
outside world. Unfair offers would provide 
the divider with a fitness advantage. 
An aversion to unfairness, which is 
functionally equivalent to envy, is 
then rational.

We specialize to $s_0=1$ and denote with 
$s\in[0,1]$ the monetary utility the 
responder would receive when accepting, 
with the proposer getting $1-s$. Applying
our generic utility (\ref{R_i_alpha})
to the ultimatum game we have
\begin{equation}
R(s) = s + \epsilon\log\big(s/\bar s\big),
\qquad\quad
\bar s = 1/2
\label{ultimatum_game}
\end{equation}
for the reward function of the responder.
We denoted with $\bar s$ the average 
monetary utility and used that the 
responder plays pure strategies 
(accepting/declining). The responder 
accepts offers leading to positive 
rewards, declining negative $R(s)$.

\begin{figure}[!t]
\centering
\includegraphics[width=0.85\columnwidth]{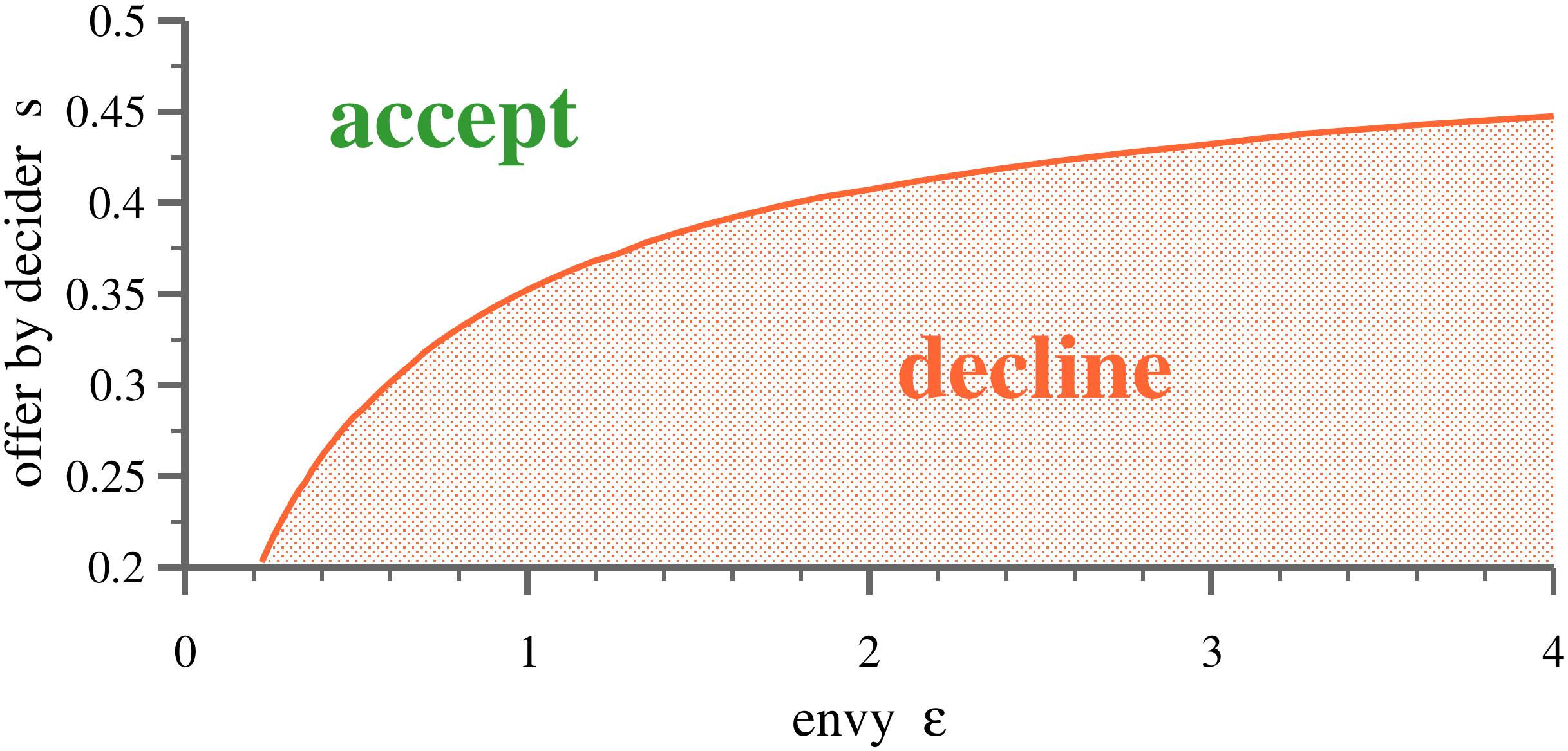}
\caption{{\bf Ultimatum game with envy.}
The decision space for the responder in the
ultimatum game with envy, as defined by
(\ref{ultimatum_game}). Even substantially unfair 
offers are accepted when the envy parameter $\epsilon$
is small, but only nearly fair offers for
large envy. Experimentally, offers below $s=0.4$
are rare \protect\cite{fehr1999theory}, which indicates
that $\epsilon \approx 1.75$.
}
\label{fig_ultimatum_game}
\end{figure}

The proposer maximizes its gains when the
offer is as unfair as possible, namely
close to $s_\epsilon$, the lowest offer 
the responder will accept. It is, in effect,
an judgment call. A large number of experiments 
have shown that $s_\epsilon \approx 0.4$
\cite{fehr1999theory}, which indicates an 
value of $\epsilon \approx 1.75$.
This value can be extracted from the
data presented in Fig.~\ref{fig_ultimatum_game},
which shows the solution of $R(s)=0$.

The order of magnitude for the strength of
envy, $\epsilon \approx 1.75$, is surprisingly
large. In comparison, the range of the monetary
utility is $s\in[0,1]$. We can hence expect,
that real-world envy parameters and monetary 
utilities are of the same order of magnitude, with 
envy being possible somewhat larger. 

Here we used the concept of envy to model
the experimental outcomes of the ultimatum
games. An alternative would be, besides others, 
to postulate that both participants dispose 
of a `self-centered inequity aversion' 
\cite{fehr1999theory}, which means that both 
players dislike unfair outcomes, to varying 
degrees, and not only the responder. In our 
case the strength of the envy parameter of the 
proposer does not enter. For the proposer, 
maximizing $1-s$ or $R(1-s)$ leads to the 
same conclusion, as both are monotonically 
increasing when $1-s>1/2$. Our view, that 
the proposer is mainly interested in 
profit maximization, draws support from 
experimental studies \cite{bearden2001ultimatum}.

The fact that responder rarely accept 
offers below 40\% implies that envy is 
impactful.  Within our framework, this
leads to a specific value of $\epsilon$, 
which is about 1.75 times the overall bare 
utility. Transferring this value to the 
case of competitive societies, we 
would have $\epsilon\approx3\times1.75
\sim 5.25$. This value arises when the 
range $v(1)-v(0)=3$ is taken for the 
reference utility. At this point it is 
important to stress that the resulting
$\epsilon$ is only a rough order 
of magnitude estimate. It suggests in
any case that envy matters and that
that our result, that values of $\epsilon\approx2.5$ 
and larger lead to a class-stratified 
state, is not just an academic exercise.

\begin{figure}[!t]
\centerline{
\includegraphics[width=0.75\columnwidth]{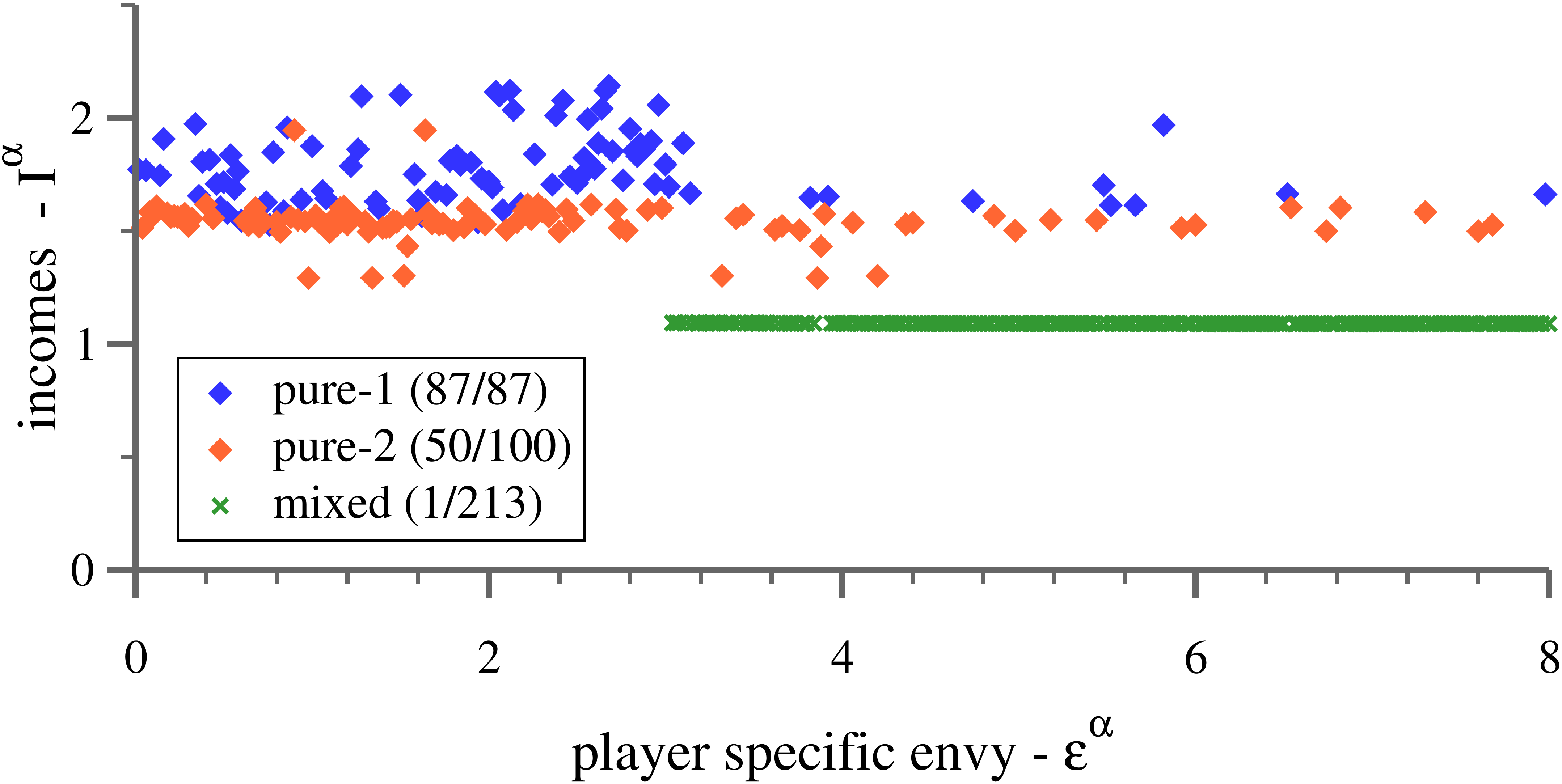}
           }
\caption{{\bf Divide-the-cake competition.}
For player specific values of envy in
the range  $\epsilon^\alpha\in[0,8]$, the 
incomes $I^\alpha$, compare 
Fig.~\ref{fig_envy_scatter_bar}.
Here we used $M=N=200$ agents and options
and the `divide-the-cake' interaction, as
specified by (\ref{I_i_alpha_cake}).
}
\label{fig_envy_scatter_cake}
\end{figure}

\subsection{Alternative formulations}

It is only rarely possible to fully analyze
non-uniform games with arbitrary large numbers
of options and agents. Examples are sealed
auctions and animal conflict models with 
continuous bidding ranges 
\cite{maskin2000equilibrium,rusch2017logic}.
The situation is more intricate for 
resource distribution games, like the 
framework examined here, as 
evidenced by the numerically obtained
structure of the payoff functions shown in
Fig.~\ref{fig_payoffs_100_rel_e10}. The
complication is in particular due to
the presence of a \'a priori not known 
number of mixed strategies, for which 
the strategy functions $p_i^\alpha$ are 
to be determined maximizing $M$ coupled
self-consistency conditions for the individual
rewards $R^\alpha$. 

An important point regards the robustness of
our scenario, namely that competitive societies
become unstable under the influence of envy.
For example, we studied the 'divide-the-cake'
type inter-agent competition
\begin{equation}
I_i^\alpha = \frac{v_i}
{1+\sum_{\beta\ne\alpha}p_i^\beta}\,,
\label{I_i_alpha_cake}
\end{equation}
which describes the situation that agents
playing pure strategies equally divide the
bare reward $v_i$ when going for one and the
same option. Only quantitative differences are
found, as illustrated in Fig.~\ref{fig_envy_scatter_cake}.

A substantially different formulation is
\begin{equation}
R_i^\alpha = 
v_i-\kappa\sum_{\beta\ne\alpha} p_i^\beta 
+\epsilon p_i^\alpha\log\left(\frac{R^\alpha}{\bar R}
\right)
\label{R_i_alpha_reward}
\end{equation}
which has been studied in the past
\cite{gros2020self,gros2021collective}.
Here $R^\alpha = \sum_i R_i^\alpha p_i^\alpha$
and $\bar R = \sum_\alpha R^\alpha/M$. Envy
is now a function of the reward, and not of
the monetary income. This implies a recursive
functionality, given that rewards depend in turn 
on the envy term. We hence believe that
that (\ref{R_i_alpha_reward}), which also
leads to a class-stratified state with
increasing $\epsilon$, is less realistic.
It has however been possible to treat the 
condensed state analytically \cite{gros2020self}.
Note that the $\kappa$-term in
(\ref{R_i_alpha_reward}) is not proportional
to $v_i$, as in (\ref{I_alpha}).

For the framework investigated here,
there is only a single references value
for comparing incomes, the population 
average $\bar{I}$. Alternatively, agents could
compare their incomes on a pairwise level, 
as discussed in the appendix. This would 
require $M-1$ operations per agent. In
between these two extremes, the average
incomes of the peer community could be
relevant, e.g.\ for societies of agents
with a well defined network structure.

\section{Conclusions\label{sect_dicsussion}}

The evolutionary drivers of core human traits, like 
cooperation \cite{sachs2004evolution,nowak2006five,west2011sixteen},
altruism \cite{trivers1971evolution,fletcher2009simple},
charity \cite{barclay2012evolution}
and revenge \cite{jackson2019revenge},
to name a few, have been investigated intensively.
To a certain extent, our propensity to value
success not just as such, but also in relative 
terms, is set apart. Given that natural selection works 
on relative success \cite{cabrales2010causes}, it
benefits evolutionary success when being sensitive 
to the success of others. From this view 
it is not surprising that non-human animals may also 
be averse to inequalities
\cite{takimoto2010capuchin,range2009absence}.
As the closest available notion we used 'envy'
to denote terms in reward functions based
on relative success. As all psychological traits, 
envy has in addition an extended palette of distinct
facets. Three key findings emerge. 

The first observation is that strategies may
condense collectively. This happens when
a large number of small mixed strategies
starts to exhaust option space, merging 
into a single big strategy. Small and big 
denote here the relative sizes of the respective 
supports, specifically when measured 
in units of overall option spaces. Small 
supports are intensive, viz not growing 
when increasing the number of options $N$, 
while keeping a constant density $M/N$ 
of agents per available options. Supports 
are in contrast big when they are extensive, 
that is when they scale with $N$. When strategies
condense, an extensive number of agents play
the identical mixed strategy, which is
characterized in turn by an extensive support.

Our second core result concerns the structure
of the society once mixed strategies did condense.
Clear gaps open in the spectrum of the monetary 
income between distinct clusters of agents. 
These gaps provide a well defined criterion
for the subdivision of the society into different
social classes. Our results lead consequently to 
the conjecture that envy may play a key role for
the original emergence of class-stratified societies.
Besides being separated by income gaps, the
three emerging clusters of agents correspond 
to distinct strategy types, with lower, 
middle and upper class agents playing 
respectively mixed, pure-2 and pure-1 
strategies, where the two types of pure 
strategies differ with respect to the
amount of competition encountered.

It is possible to generalize our framework
to the ultimatum game, which allows to
to estimate the real-world value of envy.
Comparing with the experimental results 
for the ultimatum game one finds that human 
societies are possibly located within 
the class-stratified phase.

Finally we observe that envy, as defined
here, is counterproductive for general 
welfare, which can be regarded as a 
generalized social dilemma. The average income drops
when envy becomes progressively more important,
in particular once strategy condensation sets
in. All in all we suggest that the study 
of games with mixed payoff functions, containing 
both  monetary and psychological components,
is a timely subject.

\subsection{Data Availability}

All data has been generated using 
equation (\ref{evolutionaryDynamics}).

\subsection{Acknowledgments}

We thank Carolin Roskothen for stimulating
discussions and Roser Valent\'i for comments.

\section*{Appendix}

We used in (\ref{R_i_alpha}) that
envy is based on a comparison with
the population mean. Alternatively,
as in \cite{fehr1999theory}, one could
start with a pairwise formulation for envy, 
\begin{equation}
\mathrm{Envy}\big|_\alpha= \frac{\epsilon}{M}
\sum_{\beta\ne\alpha}
\log\left(\frac{I^\alpha}{I^\beta}\right)\,,
\label{envy_def_pair}
\end{equation}
which expresses that agents make direct, one-to-one
comparisons with everybody else. As before, 
we define with 
$\bar{I} = \left(\sum_\alpha I^\alpha\right)/M$
the average monetary income. Expanding
with respect to small relative deviations
$(I^\beta-\bar{I})/\bar{I}$ from the mean, we obtain
\begin{eqnarray}
\nonumber
\log\left(\frac{I^\alpha}{I^\beta}\right)
&=& \log(I^\alpha) -\log(I^\beta) 
\\ \nonumber &=& 
\log(I^\alpha-\bar{I}+\bar{I}) -\log(I^\beta-\bar{I}+\bar{I}) 
\\ \nonumber &=& 
\log\left(1+\frac{I^\alpha-\bar{I}}{\bar{I}}\right)  -
\log\left(1+\frac{I^\beta -\bar{I}}{\bar{I}}\right) 
\\ &\approx& 
\frac{I^\alpha-I^\beta}{\bar{I}}\,.
\label{log_expansion}
\end{eqnarray}
Noting that the vanishing term $\beta=\alpha$ 
could have be added to right-hand side of 
(\ref{envy_def_pair}), we apply
the sum $(\epsilon/M)\sum_\beta$ to 
the linearized expression (\ref{log_expansion}).
The result is
\begin{equation}
\mathrm{Envy}\big|_\alpha 
\approx \epsilon\, \frac{I^\alpha-\bar{I}}{\bar{I}}\
\approx \epsilon\, \log\left(\frac{I^\alpha}{\bar{I}}\right)\,.
\label{P_alpha_linearized}
\end{equation}
which coincides in linear approximation 
with our original definition, as
used in (\ref{R_i_alpha}).


\begin{thebibliography}{10}

\bibitem{wyer2014memory}
Robert~S Wyer~Jr and Thomas~K Srull.
\newblock {\em Memory and cognition in its social context}.
\newblock Psychology Press, 2014.

\bibitem{linde2012social}
Jona Linde and Joep Sonnemans.
\newblock Social comparison and risky choices.
\newblock {\em Journal of Risk and Uncertainty}, 44(1):45--72, 2012.

\bibitem{lahno2015peer}
Amrei~M Lahno and Marta Serra-Garcia.
\newblock Peer effects in risk taking: Envy or conformity?
\newblock {\em Journal of Risk and Uncertainty}, 50(1):73--95, 2015.

\bibitem{tetlock1985accountability}
Philip~E Tetlock.
\newblock Accountability: The neglected social context of judgment and choice.
\newblock {\em Research in organizational behavior}, 7(1):297--332, 1985.

\bibitem{tindale2019group}
R~Scott Tindale and Jeremy~R Winget.
\newblock Group decision-making.
\newblock In {\em Oxford Research Encyclopedia of Psychology}. 2019.

\bibitem{milinski2002reputation}
Manfred Milinski, Dirk Semmann, and Hans-J{\"u}rgen Krambeck.
\newblock Reputation helps solve the ‘tragedy of the commons’.
\newblock {\em Nature}, 415(6870):424--426, 2002.

\bibitem{hilbe2012emergence}
Christian Hilbe and Arne Traulsen.
\newblock Emergence of responsible sanctions without second order free riders,
  antisocial punishment or spite.
\newblock {\em Scientific reports}, 2(1):1--4, 2012.

\bibitem{kurokawa2009emergence}
Shun Kurokawa and Yasuo Ihara.
\newblock Emergence of cooperation in public goods games.
\newblock {\em Proceedings of the Royal Society B: Biological Sciences},
  276(1660):1379--1384, 2009.

\bibitem{raub1997gains}
Werner Raub and Chris Snijders.
\newblock Gains, losses, and cooperation in social dilemmas and collective
  action: The effects of risk preferences.
\newblock {\em Journal of Mathematical Sociology}, 22(3):263--302, 1997.

\bibitem{hagel2016risk}
Kristin Hagel, Maria Abou~Chakra, Benedikt Bauer, and Arne Traulsen.
\newblock Which risk scenarios can drive the emergence of costly cooperation?
\newblock {\em Scientific reports}, 6(1):1--9, 2016.

\bibitem{szolnoki2016leaders}
Attila Szolnoki and Matja{\v{z}} Perc.
\newblock Leaders should not be conformists in evolutionary social dilemmas.
\newblock {\em Scientific Reports}, 6(1):1--8, 2016.

\bibitem{gershman2014two}
Boris Gershman.
\newblock The two sides of envy.
\newblock {\em Journal of Economic Growth}, 19(4):407--438, 2014.

\bibitem{alicke2008social}
Mark~D Alicke and Ethan Zell.
\newblock Social comparison and envy.
\newblock 2008.

\bibitem{hopkins2004running}
Ed~Hopkins and Tatiana Kornienko.
\newblock Running to keep in the same place: Consumer choice as a game of
  status.
\newblock {\em American Economic Review}, 94(4):1085--1107, 2004.

\bibitem{mcbride2001relative}
Michael McBride.
\newblock Relative-income effects on subjective well-being in the
  cross-section.
\newblock {\em Journal of Economic Behavior \& Organization}, 45(3):251--278,
  2001.

\bibitem{clark2010compares}
Andrew~E Clark and Claudia Senik.
\newblock Who compares to whom? the anatomy of income comparisons in europe.
\newblock {\em The Economic Journal}, 120(544):573--594, 2010.

\bibitem{gros2015complex}
Claudius Gros.
\newblock {\em Complex and Adaptive Dynamical Systems, a Primer}.
\newblock Springer, 2015.

\bibitem{hill2008evolutionary}
Sarah~E Hill and David~M Buss.
\newblock The evolutionary psychology of envy.
\newblock 2008.

\bibitem{capraro2021mathematical}
Valerio Capraro and Matja{\v{z}} Perc.
\newblock Mathematical foundations of moral preferences.
\newblock {\em Journal of the Royal Society Interface}, 18(175):20200880, 2021.

\bibitem{gros2021collective}
Claudius Gros.
\newblock Collective strategy condensation: When envy splits societies.
\newblock {\em Entropy}, 23(2):157, 2021.

\bibitem{griffin1996bose}
Allan Griffin, David~W Snoke, and Sandro Stringari.
\newblock {\em Bose-Einstein condensation}.
\newblock Cambridge University Press, 1996.

\bibitem{bianconi2001bose}
Ginestra Bianconi and Albert-L{\'a}szl{\'o} Barab{\'a}si.
\newblock Bose-einstein condensation in complex networks.
\newblock {\em Physical review letters}, 86(24):5632, 2001.

\bibitem{hecht1924visual}
Selig Hecht.
\newblock The visual discrimination of intensity and the weber-fechner law.
\newblock {\em The Journal of general physiology}, 7(2):235--267, 1924.

\bibitem{dehaene2003neural}
Stanislas Dehaene.
\newblock The neural basis of the weber--fechner law: a logarithmic mental
  number line.
\newblock {\em Trends in cognitive sciences}, 7(4):145--147, 2003.

\bibitem{howard2018memory}
Marc~W Howard.
\newblock Memory as perception of the past: Compressed time in mind and brain.
\newblock {\em Trends in cognitive sciences}, 22(2):124--136, 2018.

\bibitem{gros2012neuropsychological}
Claudius Gros, Gregor Kaczor, and Dimtrij{\'e} Markovi{\'c}.
\newblock Neuropsychological constraints to human data production on a global
  sca le.
\newblock {\em The European Physical Journal B}, 85(1):28, 2012.

\bibitem{gros2020self}
Claudius Gros.
\newblock Self induced class stratification in competitive societies of agents:
  Nash stability in the presence of envy.
\newblock {\em Royal Society Open Science}, 7:200411, 2020.

\bibitem{hofbauer2003evolutionary}
Josef Hofbauer and Karl Sigmund.
\newblock Evolutionary game dynamics.
\newblock {\em Bulletin of the American Mathematical Society}, 40(4):479--519,
  2003.

\bibitem{yakovenko2007two}
Victor~M Yakovenko and A~Christian Silva.
\newblock Two-class structure of income distribution in the usa: Exponential
  bulk and power-law tail.
\newblock In {\em Topological Aspects Of Critical Systems And Networks: (With
  CD-ROM)}, pages 49--58. World Scientific, 2007.

\bibitem{ramachandran2017evolutionary}
Vilayanur~S Ramachandran and Baland Jalal.
\newblock The evolutionary psychology of envy and jealousy.
\newblock {\em Frontiers in psychology}, 8:1619, 2017.

\bibitem{garay2011envy}
J{\'o}zsef Garay and Tam{\'a}s~F M{\'o}ri.
\newblock Is envy one of the possible evolutionary roots of charity?
\newblock {\em Biosystems}, 106(1):28--35, 2011.

\bibitem{thaler1988anomalies}
Richard~H Thaler.
\newblock Anomalies: The ultimatum game.
\newblock {\em Journal of economic perspectives}, 2(4):195--206, 1988.

\bibitem{kirchsteiger1994role}
Georg Kirchsteiger.
\newblock The role of envy in ultimatum games.
\newblock {\em Journal of economic behavior \& organization}, 25(3):373--389,
  1994.

\bibitem{fehr1999theory}
Ernst Fehr and Klaus~M Schmidt.
\newblock A theory of fairness, competition, and cooperation.
\newblock {\em The quarterly journal of economics}, 114(3):817--868, 1999.

\bibitem{bearden2001ultimatum}
Joseph~Neil Bearden.
\newblock Ultimatum bargaining experiments: The state of the art.
\newblock {\em Available at SSRN 626183}, 2001.

\bibitem{maskin2000equilibrium}
Eric Maskin and John Riley.
\newblock Equilibrium in sealed high bid auctions.
\newblock {\em The Review of Economic Studies}, 67(3):439--454, 2000.

\bibitem{rusch2017logic}
Hannes Rusch and Sergey Gavrilets.
\newblock The logic of animal intergroup conflict: a review.
\newblock {\em Journal of Economic Behavior \& Organization}, 2017.

\bibitem{sachs2004evolution}
Joel~L Sachs, Ulrich~G Mueller, Thomas~P Wilcox, and James~J Bull.
\newblock The evolution of cooperation.
\newblock {\em The Quarterly review of biology}, 79(2):135--160, 2004.

\bibitem{nowak2006five}
Martin~A Nowak.
\newblock Five rules for the evolution of cooperation.
\newblock {\em science}, 314(5805):1560--1563, 2006.

\bibitem{west2011sixteen}
Stuart~A West, Claire El~Mouden, and Andy Gardner.
\newblock Sixteen common misconceptions about the evolution of cooperation in
  humans.
\newblock {\em Evolution and human behavior}, 32(4):231--262, 2011.

\bibitem{trivers1971evolution}
Robert~L Trivers.
\newblock The evolution of reciprocal altruism.
\newblock {\em The Quarterly review of biology}, 46(1):35--57, 1971.

\bibitem{fletcher2009simple}
Jeffrey~A Fletcher and Michael Doebeli.
\newblock A simple and general explanation for the evolution of altruism.
\newblock {\em Proceedings of the Royal Society B: Biological Sciences},
  276(1654):13--19, 2009.

\bibitem{barclay2012evolution}
Pat Barclay.
\newblock The evolution of charitable behaviour and the power of reputation.
\newblock 2012.

\bibitem{jackson2019revenge}
Joshua~Conrad Jackson, Virginia~K Choi, and Michele~J Gelfand.
\newblock Revenge: A multilevel review and synthesis.
\newblock {\em Annual Review of Psychology}, 70:319--345, 2019.

\bibitem{cabrales2010causes}
Antonio Cabrales.
\newblock The causes and economic consequences of envy.
\newblock {\em SERIEs}, 1(4):371--386, 2010.

\bibitem{takimoto2010capuchin}
Ayaka Takimoto, Hika Kuroshima, and Kazuo Fujita.
\newblock Capuchin monkeys (cebus apella) are sensitive to others’ reward: an
  experimental analysis of food-choice for conspecifics.
\newblock {\em Animal cognition}, 13(2):249--261, 2010.

\bibitem{range2009absence}
Friederike Range, Lisa Horn, Zs{\'o}fia Viranyi, and Ludwig Huber.
\newblock The absence of reward induces inequity aversion in dogs.
\newblock {\em Proceedings of the National Academy of Sciences},
  106(1):340--345, 2009.

\end{thebibliography}

\end{document}